\documentclass[conference]{IEEEtran}
\IEEEoverridecommandlockouts
\usepackage{cite}
\usepackage{amsmath,amssymb,amsfonts,amsthm}
\usepackage{graphicx}
\usepackage{textcomp}
\usepackage{xcolor}
\usepackage{xkeyval}
\usepackage{wasysym}
\usepackage{times, epsfig, array, mathrsfs,multirow}
\usepackage[caption=false]{subfig}
\usepackage{stfloats}
\usepackage{threeparttable}
\usepackage{cases}
\usepackage{cite}
\usepackage[all=normal, paragraphs=tight, floats=tight, mathspacing=tight]{savetrees}

\usepackage{comment} 

\usepackage{bm}
\usepackage{algorithm}
\usepackage{algpseudocode}
\usepackage[top=0.75in, bottom=1.1in, left=0.625in, right=0.625in]{geometry}

\begin{document}
\title{Pseudo-Noise Superposition for Finite-Alphabet Physical Layer Security}

\vspace{-0.5cm}
\author{ 
\IEEEauthorblockN{Fernando Moya Caceres, Chamath Divarathne, Yapeng Xie, Michael Aygur,
\\
Sithamparanathan Kandeepan, Akram Al-Hourani, Saman Atapattu
}
 \IEEEauthorblockA{
Department of Electrical and Electronic Engineering, School of Engineering, RMIT University, Melbourne, Australia. \\
\IEEEauthorblockA{Email:
\{fernando.moyacaceres, chamath.divarathne, akram.hourani\}@rmit.edu.au
}
}
\vspace{-1.0cm}
}
\maketitle
\begin{abstract}
Physical-layer security based on pseudo-noise (PN) superposition is a promising approach for mitigating eavesdropping in future wireless systems. However, under Shannon’s capacity formulation with Gaussian signaling, achieving secrecy typically requires allocating substantial transmit power to PN, resulting in a significant reduction in achievable information rate and limiting practical applicability. This limitation is alleviated when finite-alphabet modulation schemes, such as M-ary Quadrature Amplitude Modulation (\(M\)-QAM), are employed, as expected in practical 6G transceivers. In this work, we analyze the information rate performance of PN-assisted systems under $M$-QAM signaling using mutual information and derive the corresponding achievable secrecy rate. The impact of PN power allocation on both the legitimate user and the eavesdropper is investigated across different modulation orders and channel conditions. Monte Carlo simulations are conducted to evaluate system behavior under varying user and eavesdropper channel conditions and to examine how PN power allocation influences secrecy performance. The results show that, at sufficiently high signal-to-noise ratio (SNR), the information rate becomes largely insensitive to PN power allocation, enabling near-perfect secrecy with $M$-QAM modulation—highlighting a key departure from Shannon-capacity-based secrecy analyses and underscoring the practicality of finite-alphabet security mechanisms for 6G wireless systems.
\end{abstract}


\section{Introduction}

The evolution toward sixth-generation (6G) wireless systems is expected to enable highly heterogeneous services, including massive machine-type communications and integrated terrestrial and non-terrestrial networks. In such open environments, ensuring data confidentiality and trust remains a fundamental challenge. Due to the broadcast nature of wireless channels, transmissions are inherently vulnerable to eavesdropping and jamming, which are increasingly difficult to mitigate using conventional cryptographic mechanisms alone, particularly in complexity-constrained scenarios~\cite{9733393}. This has motivated growing interest in security mechanisms that can be embedded directly into the physical layer.

Physical-layer security has emerged as a promising paradigm for enhancing confidentiality by exploiting intrinsic wireless channel characteristics, such as fading and interference, to provide information-theoretic secrecy~\cite{5580113,10552432}. By operating independently of upper-layer cryptographic protocols, physical-layer techniques are particularly attractive for 6G use cases involving partial trust relationships or untrusted infrastructure.

Among existing physical-layer security techniques, artificial noise (AN) schemes, often realised using pseudo-noise (PN) sequences, have been widely studied. In PN-aided transmission, the transmitter injects deterministic but noise-like interference to degrade the received signal quality at eavesdroppers, while legitimate receivers exploit prior knowledge of the PN sequence to recover the desired signal~\cite{1558439}. When combined with multiple antennas, PN can be spatially shaped to enhance secrecy without significantly compromising legitimate performance~\cite{4543070}.

The effectiveness of PN-based secure transmission depends on practical factors such as channel state information (CSI) availability and the power allocation between the information-bearing signal and PN~\cite{7470273}. Prior studies have shown that equal power allocation between the two components (i.e., $\zeta = 0.5$) offers a simple yet near-optimal secrecy performance in scenarios with non-colluding eavesdroppers and limited CSI, and has therefore been widely adopted as a benchmark~\cite{5524086,4543070,5701754}.

However, much of the existing literature relies on unconstrained Gaussian signalling, yielding secrecy capacity expressions based on Shannon theory. While these results provide useful upper bounds, they do not accurately reflect practical systems employing discrete modulation schemes such as M-ary Quadrature Amplitude Modulation ($M$-QAM). In realistic 6G transceivers, secrecy performance is more appropriately characterised using constellation-constrained mutual information (finite-alphabet inputs), which reflects the achievable secrecy rate under the deployed discrete modulation rather than an idealized Gaussian input~\cite{6310164}.

In this paper, we study a downlink communication scenario in which a transmitter serves a legitimate user in the presence of eavesdroppers. PN signal is superposed with an $M$-QAM modulated information signal to intentionally degrade the reception quality at unintended receivers. The resulting physical-layer security enhancement and the associated power allocation between the information and PN signals are analysed. The main contributions are summarised as follows:
\begin{itemize}
    \item We analyse the secrecy performance of PN-aided transmission under $M$-QAM signalling and compare it with the upper-bound given by Shannon’s secrecy capacity.
    \item Mathematical expressions are derived for the achievable secrecy rate based on constellation-constrained mutual information.
    \item An optimal power allocation strategy between the information signal and PN is characterised as a function of user signal-to-noise ratio (SNR) and the channel quality ratio.
    \item A secrecy-oriented adaptive $M$-QAM modulation scheme is proposed to maximise the achievable secrecy rate.
\end{itemize}

\section{System Model}\label{Sec_SysModel}
\begin{figure}[t!]
    \centering
    \includegraphics[width=0.48\textwidth]{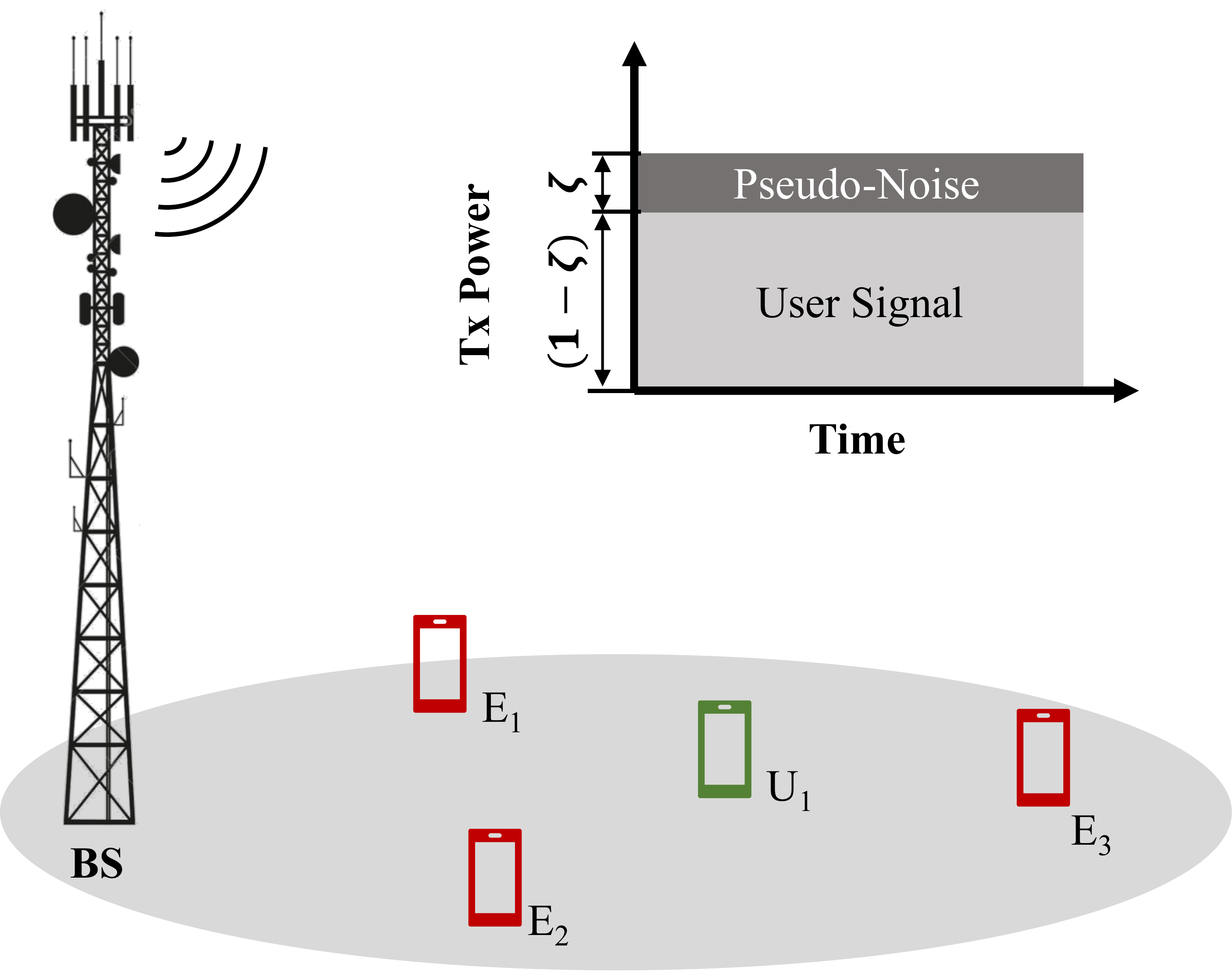}
    \caption{System Model: Downlink transmission using a superposition of user signal and pseudo-noise signal in the presence of eavesdroppers.}
    \label{fig1:System_Model}
\end{figure}
Consider the downlink communication system depicted in Fig. \ref{fig1:System_Model}, where a Base Station (BS) serving a single-antenna user terminal surrounded by multiple single-antenna eavesdropping terminals attempting to overhear the communication link. To address this issue and mitigate eavesdropping on the communication link, we superpose a PN signal at the transmitter side with the user signal by allocating a portion of the available transmit power at the BS to each of the PN and user signals. This has the purpose of creating an interfering signal on any unauthorized terminal, hindering the signal decoding process. Assume a unit total transmit power, where $\zeta$ denotes the fraction of power allocated to the PN signal and $1-\zeta$ represents the fraction assigned to the user signal.

\begin{figure}[htb]
    \centering
    \subfloat[]{
        
        \includegraphics[width=0.48\columnwidth]{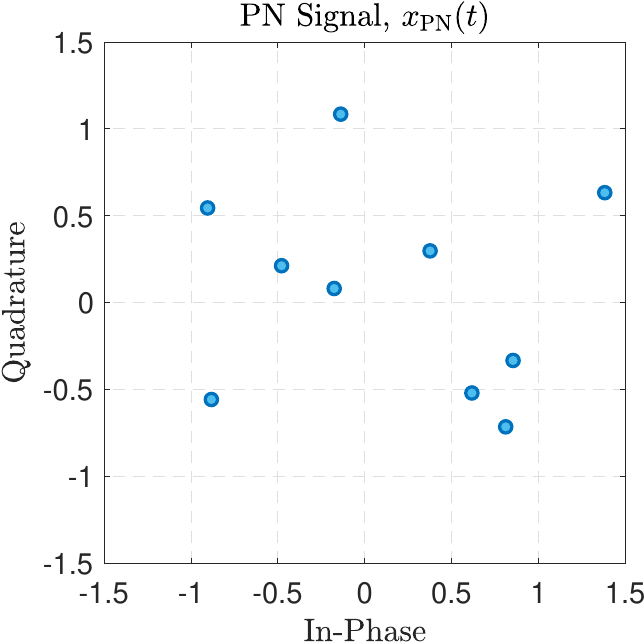}
        \label{fig:IQ_PN}
      
    }
    \subfloat[]{
        
        \includegraphics[width=0.48\columnwidth]{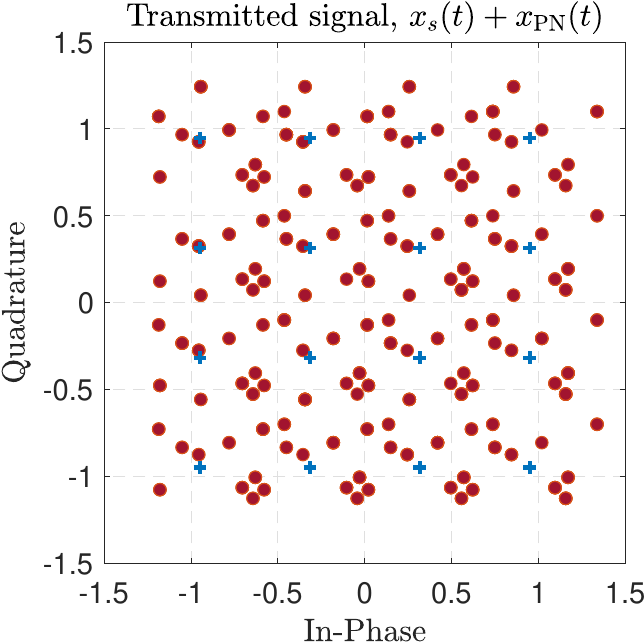}
        \label{fig:IQ_User+PN}
    }
    \vspace{1mm}
    \subfloat[]{
        
        \includegraphics[width=0.48\columnwidth]{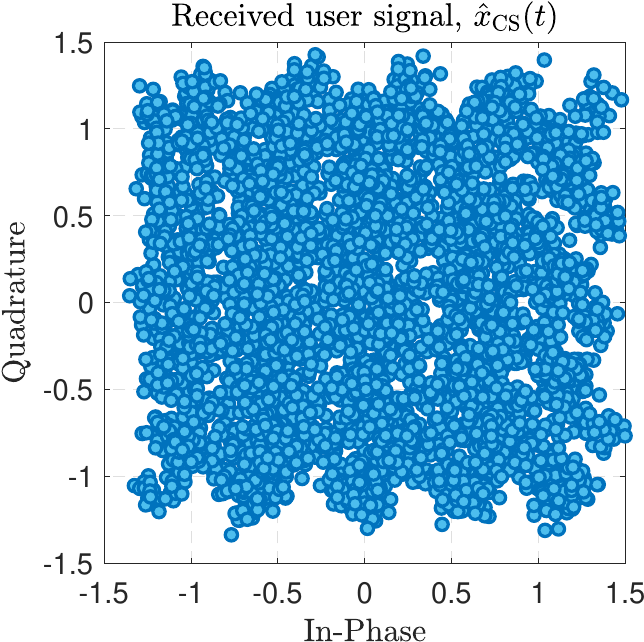}
        \label{fig:IQ_PN_receiver}
      
    }
    \subfloat[]{
        
        \includegraphics[width=0.48\columnwidth]{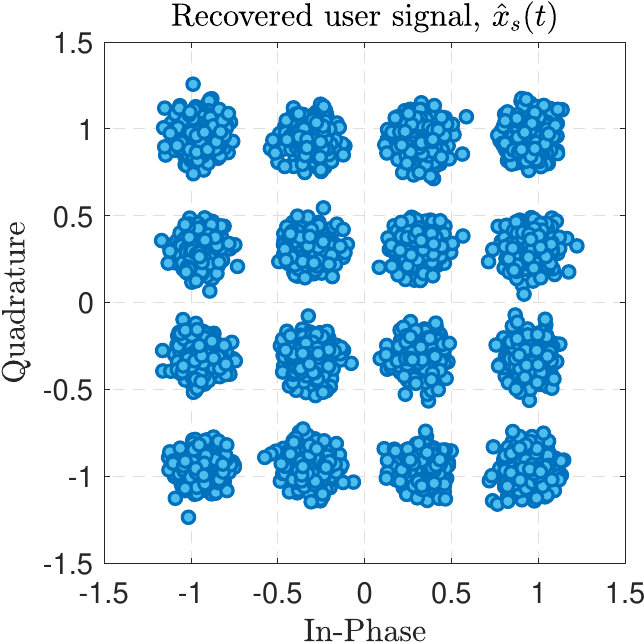}
        \label{fig:IQ_receiver}
    }
    \caption{Signal Constellation. (a) PN signal (\(\zeta=0.05\)), (b) Transmitted signal (16-QAM) + PN signal (\(\zeta=0.05\)), (c) Received composite signal and (d) Recovered signal.} 
    \label{fig:Constellation_Diagram}
\end{figure}
\begin{table}
\caption{\textbf{Notations Used}}
\setlength{\tabcolsep}{3pt}
\begin{tabular}{|p{25pt}|p{200pt}|}
\hline
Symbol&
Explanation \\
\hline
$x_\mathrm{s}(t)$&
User data signal \\
$x_\mathrm{PN}(t)$&
Pseudo-noise signal \\
$\hat{x}_\mathrm{CS}(t)$&
Estimated composite signal of user data and pseudo-noise \\
$n(t)$&
Noise process at the receiver\\
$y_u(t)$&
Received signal at user \\
$y_e(t)$&
Received signal at eavesdropper \\
$h_u$&
User channel \\
$h_e$&
Eavesdropper channel \\
$\zeta$&
Power allocation ratio for pseudo-noise signal \\
\hline
\end{tabular}
\label{Table:Notation}
\end{table}
Considering the notation in Table \ref{Table:Notation}, the received signal at user \(u\) or at eavesdropper \(e\) is given by 
\begin{equation}
\label{eq:Received_Signal_User_Eavesdropper}
y_i(t) = \underbrace{\left(\sqrt{1-\zeta}\right) h_i x_s(t)}_{\text{User Signal}}+\underbrace{\sqrt{\zeta} h_i x_\mathrm{PN}(t)}_{\text{PN signal}}+n_i(t),
\end{equation}
where \(i\in\{u,e\}\). Notice that \({n}_i(t)\sim \mathbb{C}\mathcal{N}(0,\sigma_i^2)\) is the complex additive white Gaussian noise (AWGN) with zero mean and variance \(\sigma_i^2\). Without loss of generality, we consider \(x_s(t)\) to have unit average energy \(\left(\mathbb{E}\left[|x_s(t)|^2\right]=1\right)\). Similar considerations are applied to the pseudo-noise signal \(x_\mathrm{PN}(t)\) which is deterministic but possesses noise-like characteristics, including a broad spectral distribution and favorable autocorrelation properties. The sequence is generated from a shared secret seed known only to legitimate nodes. By employing identical pseudo-random number generators and maintaining synchronization, the transmitter and legitimate receivers independently reconstruct the same PN sequence for insertion and cancellation, respectively.. To characterise a worst-case secrecy scenario, both the legitimate user and the eavesdropper are assumed to have access to ideal CSI. For analytical clarity, and without loss of generality, the channel gain is normalised to unity. Under this assumption, the composite signal comprising the user data and pseudo-noise, $\hat{x}_{\mathrm{CS}}(t)$, is obtained at both the legitimate receiver and the eavesdropper. The knowledge of the PN signal at the legitimate user enables local reconstruction and PN mitigation. Hence, the estimated user data \(\left(\hat{x}_s(t)\right)\) is calculated using the known \(x_\mathrm{PN}(t)\) and \(\zeta\) to mitigate the interfering PN signal as shown in Eq. \eqref{eq4}.
\begin{equation}
\label{eq4}
    \hat{x}_s(t)=\frac{\hat{x}_\mathrm{CS}(t)-\sqrt{\zeta} x_\mathrm{PN}(t)}{\sqrt{1-\zeta}}
\end{equation}
Fig. \ref{fig:Constellation_Diagram} illustrates the constellation diagrams observed at different stages of the proposed communication system. Notice that the parameters have been selected to constitute a representative example only. Fig. \ref{fig:IQ_PN} shows the constellation of the PN signal. In  Fig. \ref{fig:IQ_User+PN} we have the composite signal detailed before and also includes the original constellation points for the user signal for reference in red. Fig. \ref{fig:IQ_PN_receiver} illustrates the composite signal at the receiver and Fig. \ref{fig:IQ_receiver} shows the user signal at the receiver once PN has been removed. The SNR at the user terminal can be calculated as,
\begin{equation}
\gamma_{u}=\frac{(1-\zeta)|h_u|^2}{\sigma^2_u}.
    \label{Eq:gamma_user}
\end{equation}
Unlike the legitimate user, the eavesdropper is unaware of $x_\mathrm{PN}(t)$ and $\zeta$, hence will not be able to recover the true user data signal, $x_\mathrm{s}(t)$. Instead, it attempts to decode the composite signal $\hat{x}_\mathrm{CS}(t)$ which is already interfered by PN. From the eavesdropper’s perspective, the PN signal is deterministic but unknown, and is thus treated as Gaussian noise. This constitutes a worst-case assumption and results in a conservative secrecy rate analysis.

Similar to (\ref{Eq:gamma_user}), the effective SINR at the eavesdropping terminal is calculated as
\begin{equation}
\gamma_{e}=\frac{(1-\zeta)|h_e|^2}{\zeta|h_e|^2+\sigma^2_e}.
    \label{Eq:gamma_eaves}
\end{equation}

\subsection{Secrecy Rate Definitions}\label{SubSec:Secrecy_Rate_Definitions}
Let $R(\cdot)$ denote a rate-mapping function that transforms the received $\mathrm{SNR}$/$\mathrm{SINR}$ into an information rate. In this work, $R(\cdot)$ is instantiated either as the constellation-constrained achievable rate, i.e., the mutual information $I(X;Y)$ for a given $M$-QAM constellation, or as the Shannon capacity $C(\gamma_i)=\log_2(1+\gamma_i)$, which serves as an upper bound under ideal transmission assumptions, including infinite block length, optimal channel coding, accurate CSI, and transmitted symbols drawn from a continuous Gaussian distribution \cite{6773024}.

The level of protection that a system can guarantee is quantified using the secrecy rate (\(\mathrm{SR}\)) performance indicator, defined below
\begin{equation}
    \scalebox{0.91}{$
    \mathrm{SR}_u\left(\gamma_u,\gamma_e\right)=\max\left[0,R\left(\gamma_u\right)-R\left(\gamma_e\right)\right] \triangleq \left[R\left(\gamma_u\right)-R\left(\gamma_e\right) \right]^+$}.
    \label{eq:secrecy_capacity}
\end{equation}
It represents the maximum rate at which information to user \(u\) can be securely transmitted without being overheard by the eavesdropping terminal \(e\). It is defined as the positive difference between the achievable information rate \(R\) at the user terminal and at the eavesdropper terminal. 
\begin{figure}[t!]
    \centering
    \includegraphics[width=0.48\textwidth]{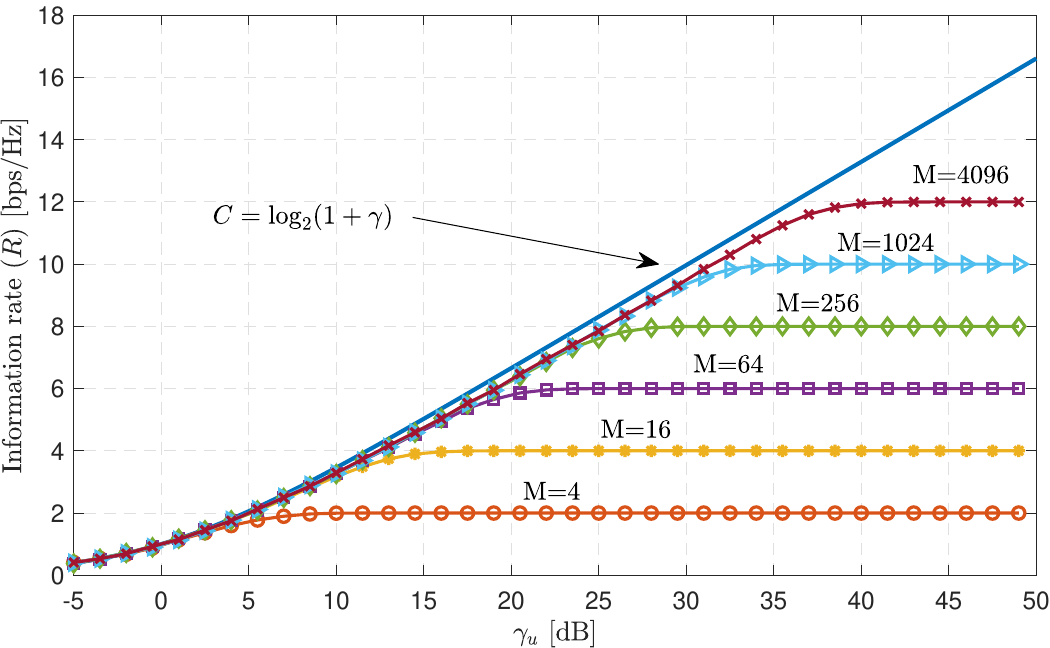}
    \caption{Comparison of achievable information rates between Shannon capacity and constellation-constrained rates for $M$-QAM as a function of user SNR.} 
    \label{fig:M-QAM_ChannelCapacity}
\end{figure}
 When \(C\) is considered for SR calculation, it is referred to as secrecy capacity (SC). However, as shown in Fig. \ref{fig:M-QAM_ChannelCapacity}, practical communication systems rely on finite-order modulation schemes, thereby limiting the achievable information rate based on the selected modulation order. At very low SNR, where noise dominates, the distribution mismatch introduced by finite-order modulation has a negligible effect, and the achievable rate closely follows Shannon’s limit. As SNR increases, finite constellation constraints become increasingly significant, resulting in a small but growing gap. In the high-SNR regime, the achievable rate ultimately saturates at $\log_2(M)$, reflecting the fundamental limit imposed by the finite modulation order on the information bits conveyed per symbol. 
  \subsection{$M$-QAM Secrecy Rate}\label{SubSec:Secrecy_Rate_M-QAM}
 For a constellation-constrained communication system (e.g., uniform $M$-QAM), the achievable information rate can be quantified using the mutual information $I(X;Y)$ (in bits/channel use) \cite{Rajski1961,6773024}, i.e., $R = I(X;Y)$. $I$ is a fundamental concept in communications and information theory that quantifies the statistical dependence between the transmitted random variable $X \in \mathcal{X}=\{x_1,\dots,x_M\}$ and the received random variable $Y$. Consistent with the system model introduced earlier, and assuming a unit channel gain without loss of generality, the received random variable is given by $Y = X + N$, where $N \sim \mathcal{CN}(0,\sigma^2)$, and the mutual information is defined as in \eqref{eq:MutualInformation}.
\begin{equation}
    I(X;Y) = \mathbb{E} \left[ \log_2 \left( \frac{p_{Y|X}(Y|X)}{p_Y(Y)} \right) \right].
    \label{eq:MutualInformation}
\end{equation}
It is defined as the expectation, with respect to the joint distribution of $X$ and $Y$, of the information density $\log_2\!\left(\frac{p_{Y|X}(y|x)}{p_Y(y)}\right)$.
Assuming a memoryless channel, the joint distribution factorises as $p_{X,Y}(x,y) = p_X(x)\,p_{Y|X}(y|x)$. Furthermore, for a constellation-constrained input such as $M$-QAM, where $X$ takes values from a finite alphabet $\mathcal{X}=\{x_1,\dots,x_M\}$, \eqref{eq:MutualInformation} can be rewritten as
\begin{equation}
\scalebox{0.97}{$
    I(X;Y) =\! \sum_{i=1}^{M} P(X=x_i)\! \int_{\mathbb{C}} p_{Y|X}(y|x_i) \log_2 \left( \frac{p_{Y|X}(y|x_i)}{p_Y(y)} \right) dy.
    $}
    \label{eq:MutualInformationEqv}
\end{equation}
To maximize the entropy and achieve the maximum achievable information rate for the given constellation, we consider that all transmitted symbols are equally probable, as \(P(X=x_i)=\frac{1}{M}, \forall x_i\in\mathcal{X}\), and knowing that \(p_{Y|X}(y|x)\) is given as
\begin{equation}
p_{Y|X}(y|x_i) = \frac{1}{\pi \sigma^2} \exp\left( -\frac{|y - x_i|^2}{\sigma^2} \right),\label{eq:Y_conditionalProb}
\end{equation}
the marginal probability density of \(Y\) is obtained following the law of total probability as \(p_Y(y) = \sum_{k=1}^{M} P(X=x_k)p_{Y|X}(y|x_k)\) and written below
\begin{equation}
p_Y(y) = \frac{1}{M} \sum_{k=1}^{M} \frac{1}{\pi \sigma^2} \exp\left( -\frac{|y - x_k|^2}{\sigma^2} \right).
\label{eq:marginalP_Y}
\end{equation}
Substituting \eqref{eq:Y_conditionalProb} and \eqref{eq:marginalP_Y} into \eqref{eq:MutualInformationEqv}, and after standard algebraic manipulations, we obtain

\begin{equation}
\begin{split}
    I(X;Y) =& \log_2 M - \frac{1}{M} \sum_{i=1}^M \int_{\mathbb{C}} \frac{1}{\pi \sigma^2} \exp\left(-\frac{|Y - x_i|^2}{\sigma^2} \right)\\
    &\log_2 \left( \sum_{k=1}^M \exp\left( -\frac{|Y - x_k|^2 - |Y - x_i|^2}{\sigma^2} \right) \right) dy,
\end{split}
\label{eq:MutualInformationEqv2}
\end{equation}
which can be conveniently expressed as
\begin{equation}
\begin{aligned}
I(X;Y)
&= \log_2 M 
- \frac{1}{M} \sum_{i=1}^M
\mathbb{E}_{Y|X=x_i}
\!\left[
\log_2
\sum_{k=1}^M
e^{-\frac{\Delta_{ik}(Y)}{\sigma^2}}
\right],
\end{aligned}
\label{eq:MutualInformationEqv3}
\end{equation}
where \(\Delta_{ik}(Y)
\triangleq |Y-x_k|^2 - |Y-x_i|^2\).
In contrast to the unconstrained Gaussian-input case associated with Shannon capacity, the mutual information for finite-alphabet modulation (e.g., $M$-QAM) does not admit a simple closed-form expression, and is typically evaluated numerically. To obtain the secrecy rate expression for a $M$-QAM modulation scheme, we write \eqref{eq:MutualInformationEqv3} in terms of \(\gamma_u\) and \(\gamma_e\) as follows
\begin{equation}
\begin{aligned}
\mathrm{SR}^{M\mathrm{\text{-}QAM}}
&= \bigl[I_u(\gamma_u) - I_e(\gamma_e)\bigr]^+ \\
&= \frac{1}{M}\Bigg[
 \sum_{i=1}^M
\mathbb{E}_{Y|X=x_i}
\left[
\log_2
\sum_{k=1}^M
\exp\!\left(
-\alpha_e \, \Delta_{ik}(Y)
\right)
\right] \\
&\hphantom{= \Bigg[}
-
\sum_{i=1}^M
\mathbb{E}_{Y|X=x_i}
\left[
\log_2
\sum_{k=1}^M
\exp\!\left(
-\alpha_u \, \Delta_{ik}(Y)
\right)
\right]
\Bigg]^+,
\end{aligned}
\label{eq:SC_M-QAM1}
\end{equation}
where \(\alpha_e\) and \(\alpha_u\) are defined as
\begin{equation}
\alpha_e
\triangleq
\frac{\gamma_e}
{|h_e|^2(1-\zeta-\gamma_e\zeta)},
\qquad
\alpha_u
\triangleq
\frac{\gamma_u}
{|h_u|^2(1-\zeta)} .
\label{eq:AlphaDef}
\end{equation}
Consequently, the secrecy rate ($\mathrm{SR}^{M\mathrm{-QAM}}$) expression in \eqref{eq:SC_M-QAM1} inherits the non-closed form nature of the $M$-QAM mutual information expressions.
\subsection{Power Allocation Ratio Selection}\label{SubSec:Power_Selection}
A central objective of this work is to determine the power allocation ratio that maximizes the secrecy rate for a given user SNR, $\gamma_u$ and modulation order, $M$. Conventional adaptive modulation techniques are first employed to select the modulation order, $M$ appropriate for the given $\gamma_u$. With a focus on maximizing secrecy performance rather than user capacity alone, the mutual information of both the legitimate user and the eavesdropper is then evaluated. Following established approaches in finite-alphabet information theory \cite{6156473, 9263098}, the mutual information is evaluated numerically, and the pseudo-noise (PN) power allocation ratio, denoted by \((\zeta)\), is optimized based on these evaluations. This procedure yields \(\zeta^\star\), the optimal PN power allocation ratio that maximizes the achievable secrecy rate for a given $M$. 
By adopting this approach, we capture the inherent complexity of secrecy analysis under practical modulation constraints while ensuring that the results remain directly applicable to real-world communication systems.

\section{Numerical Results}\label{Sec_Results}
Monte Carlo simulations are conducted to evaluate the secrecy rate performance in a scenario where an eavesdropper wiretaps the communication link between the BS and the legitimate user. As a starting point, we consider a baseline scenario in which the legitimate receiver and the eavesdropper are assumed to be co-located and therefore experience identical channel conditions. This simplifying assumption is subsequently relaxed to evaluate the impact of differing channel conditions on the user and on the eavesdropper. 
\begin{figure}[htb]
    \centering
    \subfloat{
        \includegraphics[width=0.98\columnwidth]{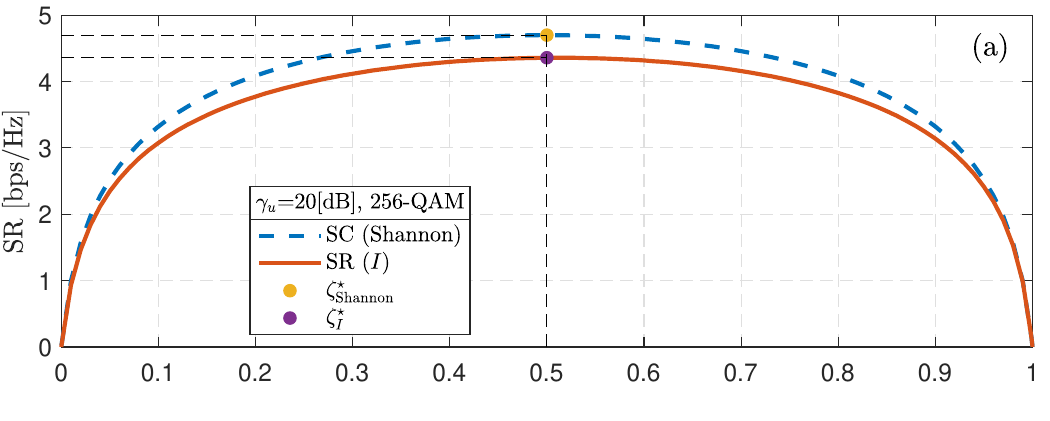}
        \label{fig:suba_64QAM_AN_10dB}
        
    }
    \vspace{1mm}
    \subfloat{
        \includegraphics[width=0.98\columnwidth]{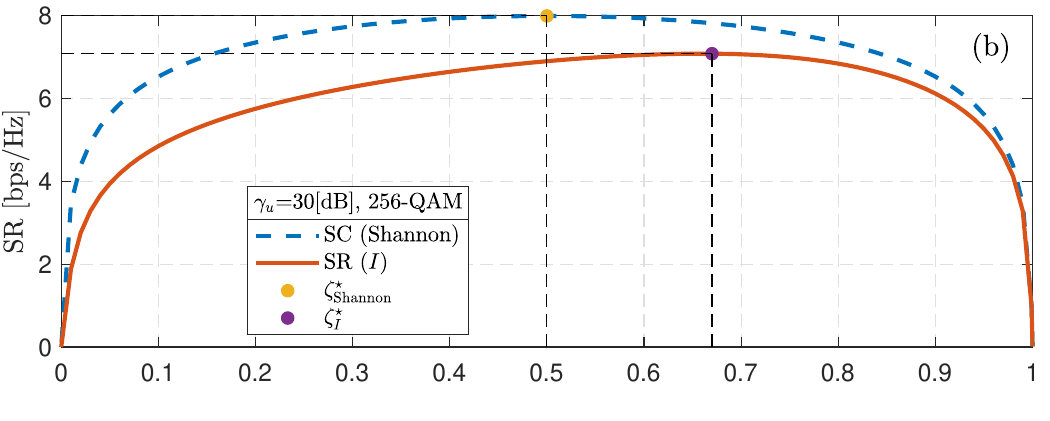}
        \label{fig:subb_64QAM_AN_24dB}
    }
    \vspace{1mm}
    \subfloat{
        \includegraphics[width=0.98\columnwidth]{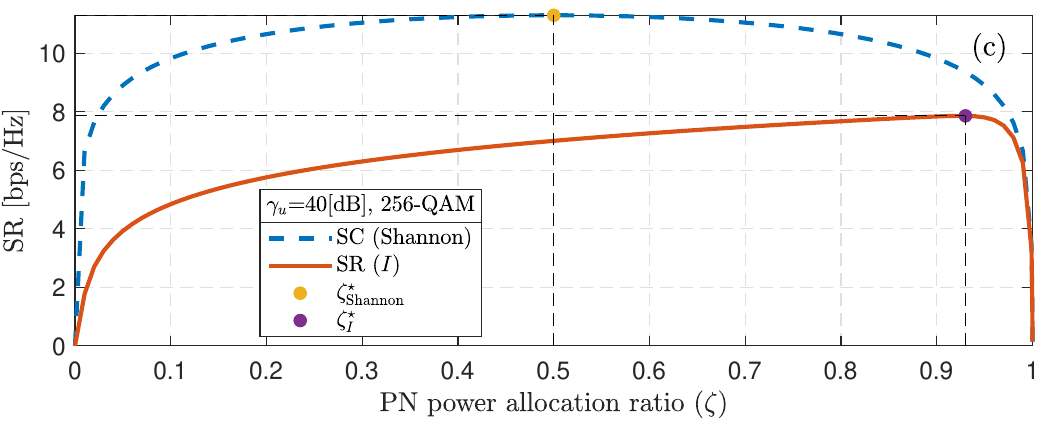}
        \label{fig:subc_64QAM_AN_30dB}
    }
    \caption{Shannon's secrecy capacity and mutual information-based secrecy rate comparison versus PN power allocation ratio \((\zeta)\) considering 256-QAM modulation order. User and eavesdropper are collocated \(\left(|h_u|^2=|h_e|^2\right)\). (a) \(\gamma_u=20 [\mathrm{dB}]\), (b) \(\gamma_u=30 [\mathrm{dB}]\) and (c) \(\gamma_u=40 [\mathrm{dB}]\).}
    \label{fig:SC_ST_256QAM_many_SNR}
\end{figure}
Fig. \ref{fig:SC_ST_256QAM_many_SNR} compares the secrecy capacity with the mutual-information-based secrecy rate for 256-QAM user data transmission at three different user SNR values (20, 30 and 40~[dB]). For all the cases considered, we observe that the optimal PN power allocation ratio (\(\zeta\)) for finite-alphabet signaling satisfies \(\zeta^\star_I\geq\zeta^\star_\mathrm{Shannon}\). Moreover, the gap between these two optimal power allocation factors increases with the user SNR. We further verify that \(\zeta^\star_\mathrm{Shannon}=0.5\), independent of variations in the user \(\gamma\). Another important observation from the three sub-figures is that the discrepancy between the 256-QAM mutual-information-based secrecy rate and the secrecy capacity becomes more pronounced at higher SNR levels. This behavior stems from the inherent information rate limitation imposed by the finite-alphabet modulation scheme, in contrast to the unbounded growth predicted by the theoretical Shannon capacity.
Continuing with the collocated legitimate user and eavesdropper scenario, Fig. \ref{fig:Opt_zeta_vs_SNR} depicts the optimal power allocation factor \(\zeta^\star\) as a function of the user SNR for four different $M$-QAM modulation orders. We confirm that the secrecy capacity based on Shannon's formulation is maximized at \(\zeta^\star_\mathrm{Shannon}=0.5\), regardless of the user's SNR, highlighting a fundamental difference compared with finite-alphabet signaling. In contrast, the optimal power allocation factor obtained from mutual-information-based secrecy rate maximization, denoted \(\zeta^\star_I\), varies with the user SNR, and each curve exhibits distinct behavior depending on the modulation order. This arises from the inherent information rate limitation of $M$-QAM signaling, which is reached at different SNR thresholds for each modulation order. Once the user's SNR exceeds this threshold additional transmit power can be allocated for PN signal without degrading the user's information rate. This in turn, enhances the secrecy performance, enabling near-perfect secrecy levels in the high SNR regime. 
\begin{figure}[htb]
    \centering
    \includegraphics[width=0.47\textwidth]{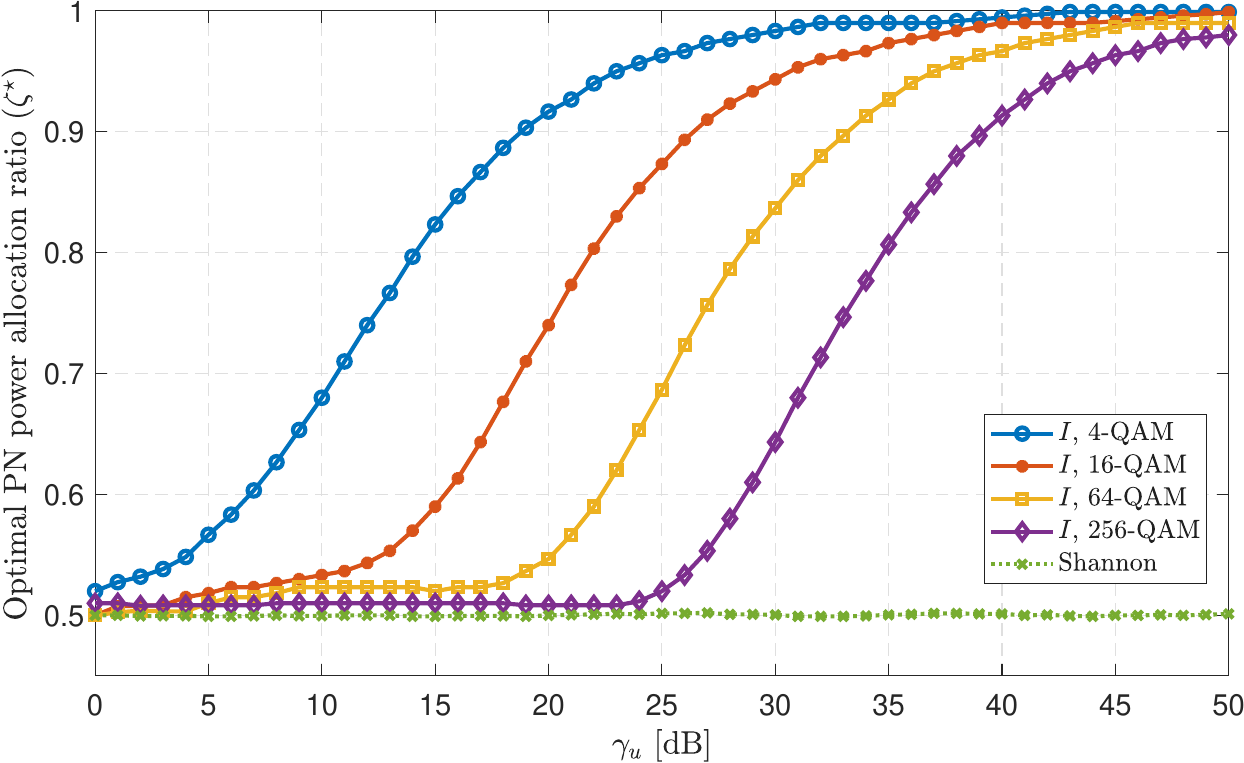}
    \caption{Optimal PN power allocation ratio (\(\zeta^\star\)) versus user SNR for 4, 16, 64 and 256-QAM modulation orders as function of user SNR. User and eavesdropper are collocated \(\left(|h_u|^2=|h_e|^2\right)\).}
    \label{fig:Opt_zeta_vs_SNR}
\end{figure}
Fig. \ref{fig:Opt_SC_vs_SNR} illustrates the maximum achievable secrecy rate for different $M$-QAM modulation orders as a function of the user SNR. For reference, the secrecy capacity curve is also included for comparison. From the figure, one can identify the user SNR threshold required for each modulation order to achieve its maximum secrecy rate. Notably, this maximum secrecy rate coincides with the maximum achievable information rate of the corresponding modulation scheme in the high SNR regime, indicating that near-perfect secrecy is attainable under finite-alphabet signaling.
\begin{figure}[htb]
    \centering
   \vspace{2mm}
\includegraphics[width=0.47\textwidth]{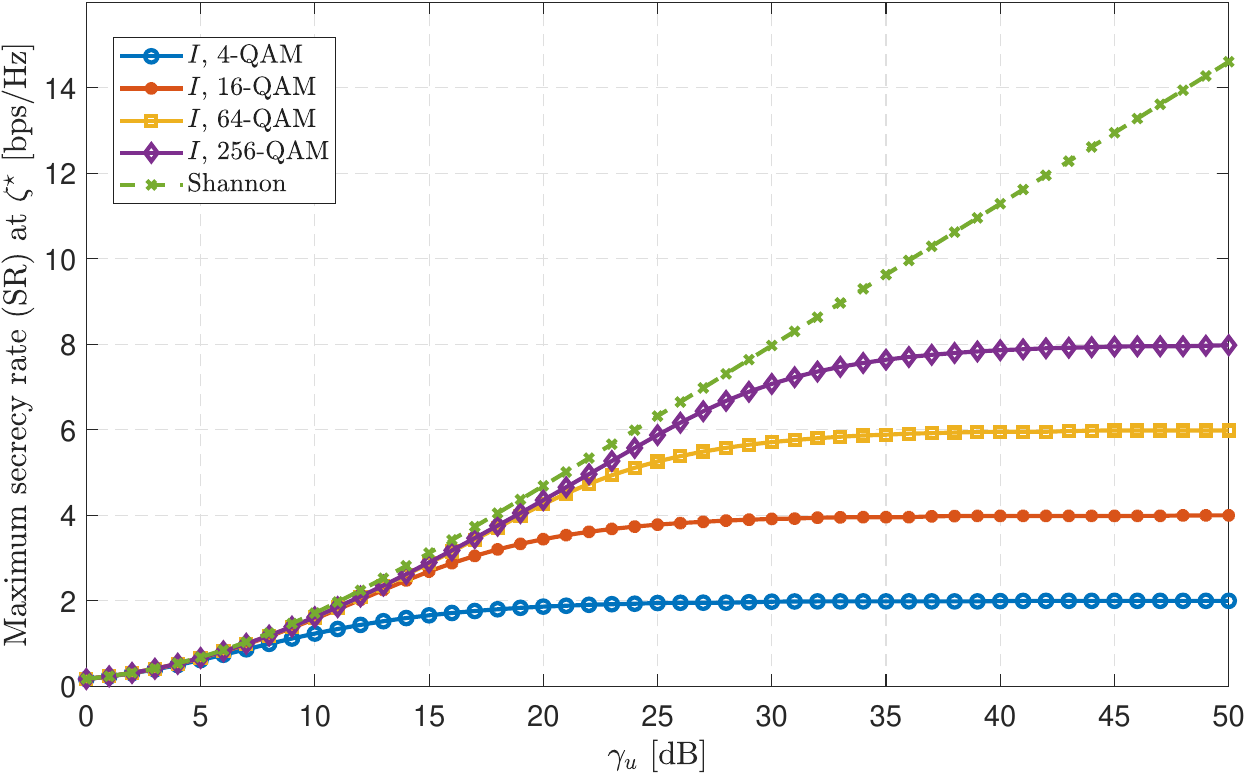}
    \caption{Maximum secrecy rate at \(\zeta^\star\) versus user SNR for 4,16, 64 and 256-QAM modulation orders as function of user SNR. User and eavesdropper are collocated \(\left(|h_u|^2=|h_e|^2\right)\).} 
    \label{fig:Opt_SC_vs_SNR}
\end{figure}
Fig. \ref{fig:Opt_overheard_ratio_vs_SNR} compares the maximum overheard information ratio, denoted by \(\eta=1-\frac{SR(\zeta^\star)}{I(\zeta^\star)}\), for various $M$-QAM modulation orders and for Shannon's capacity formulation across different SNR levels. The metric represents the maximum fraction of information intended for the legitimate user that remains unprotected and can therefore be intercepted by the eavesdropper. All considered modulation orders are capable of achieving near-zero information leakage provided that a sufficiently high SNR is available. Nevertheless, the SNR region in which high secrecy performance is attained differs across modulation orders and is closely related to the size of the modulation alphabet. In contrast, Shannon's capacity cannot guarantee perfect secrecy, since the capacity grows limitlessly with SNR. Finally, lower modulation orders require relatively smaller SNR values to achieve near-perfect secrecy.
\begin{figure}[htb]
\centering
\includegraphics[width=0.47\textwidth]{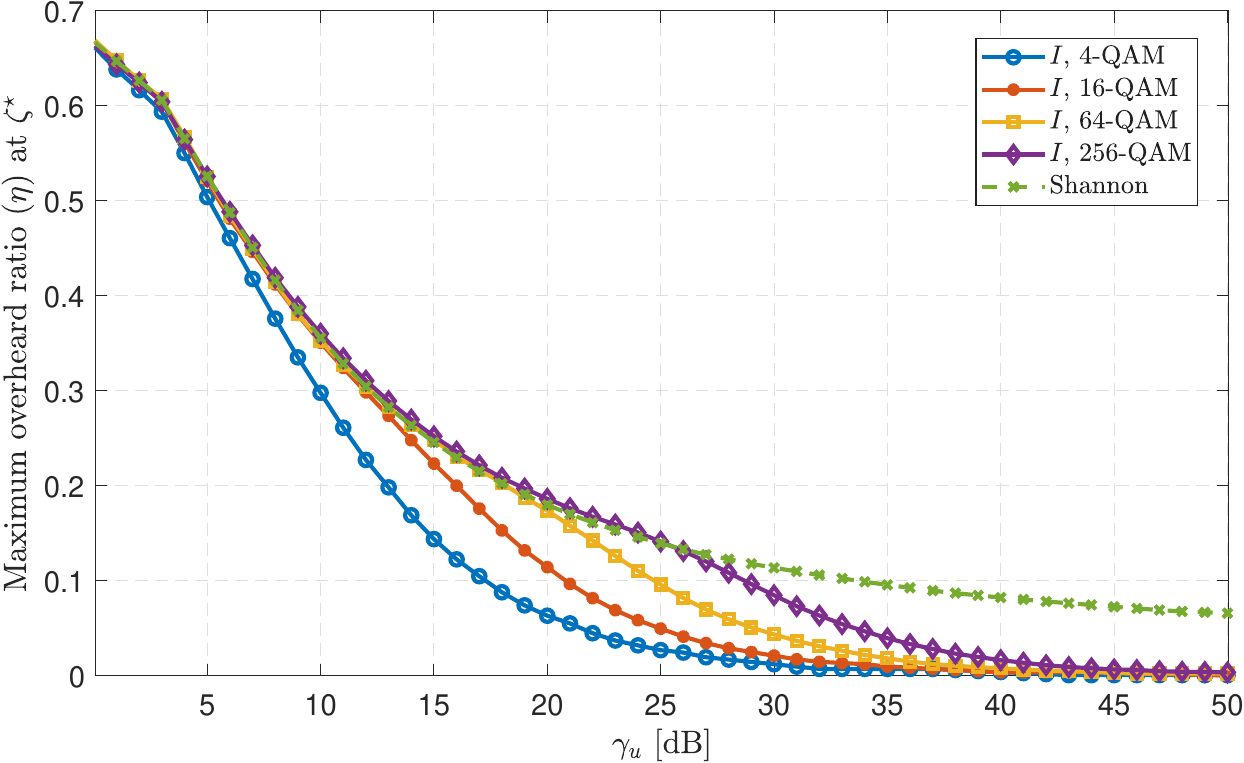}
\caption{Maximum overheard ratio \((\eta\)) by eavesdropper at optimal PN power allocation ratio \(\zeta^\star\) versus user SNR for 4,16, 64 and 256-QAM modulation orders. User and eavesdropper are collocated \(\left(|h_u|^2=|h_e|^2\right)\).} 
\label{fig:Opt_overheard_ratio_vs_SNR}
\end{figure}
In addition to the fair case scenario, we also evaluated the secrecy rate performance under asymmetric channel conditions, where the legitimate user and the eavesdropper experience different channel gains as illustrated in Fig. \ref{fig1:System_Model}. This was achieved by varying the eavesdropper-to-user channel gain ratio \(\frac{|h_e|^2}{|h_u|^2}\), considering both \(|h_e|^2>|h_u|^2\) and \(|h_e|^2<|h_u|^2\). Figures \ref{fig:Opt_zeta_vs_h_ratio_SNRs} and \ref{fig:Opt_zeta_vs_h_ratio_20dB_M_QAM} depict the optimal PN power allocation factor for this scenario. 
\begin{figure}[t]
    \centering
    \includegraphics[width=0.47\textwidth]{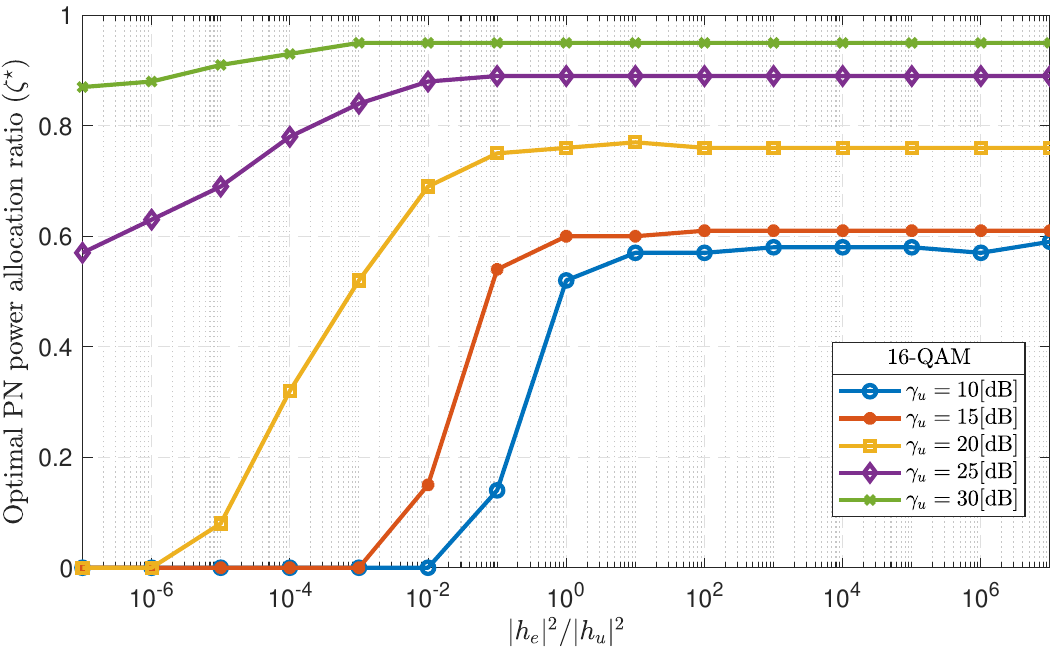}
    \caption{Optimal PN power allocation ratio (\(\zeta^\star\)) vs \(|h_e|^2/|h_u|^2\) for 16-QAM modulation order across different user SNR levels.}
    \label{fig:Opt_zeta_vs_h_ratio_SNRs}
\end{figure}
Fig. \ref{fig:Opt_zeta_vs_h_ratio_SNRs} presents results for 16-QAM at five different SNR levels. We observe that \(\zeta^\star_I\) increases with the channel gain ratio \(\frac{|h_e|^2}{|h_u|^2}\) and converges to a SNR-dependent constant value. This behavior reflects the fact that poor channel conditions at the eavesdropper substantially reduce the need to allocate transmit power to the artificial noise signal. Conversely, when the eavesdropper’s channel becomes stronger than that of the legitimate user, the optimal power allocation tends to a steady value, indicating a saturation in the required artificial noise power to maintain secrecy.  
In Fig. \ref{fig:Opt_zeta_vs_h_ratio_20dB_M_QAM} we study the optimal PN ratio behavior at a fixed SNR level \(\left(\gamma_u=20[\mathrm{dB}]\right)\)under 4 different modulation orders. Additionally, Shannon's behavior is also included for comparison purposes. Each modulation curve follows different \(\zeta^\star_I\) at similar \(\frac{|h_e|^2}{|h_u|^2}\) ratios. The fewer the number of symbols in the chosen modulation order alphabet, the higher \(\zeta^\star_I\) needs to be to maximize the secrecy rate. This is because more noise presence is needed to confuse the eavesdropper receiver under low order modulation schemes. As the modulation order increases (256-QAM case), the $M$-QAM curve closely resembles Shannon's secrecy capacity and \(\zeta^\star_I\) closely follows \(\zeta^\star_{\mathrm{Shannon}}\).
\begin{figure}[htb]
    \centering
    \includegraphics[width=0.47\textwidth]{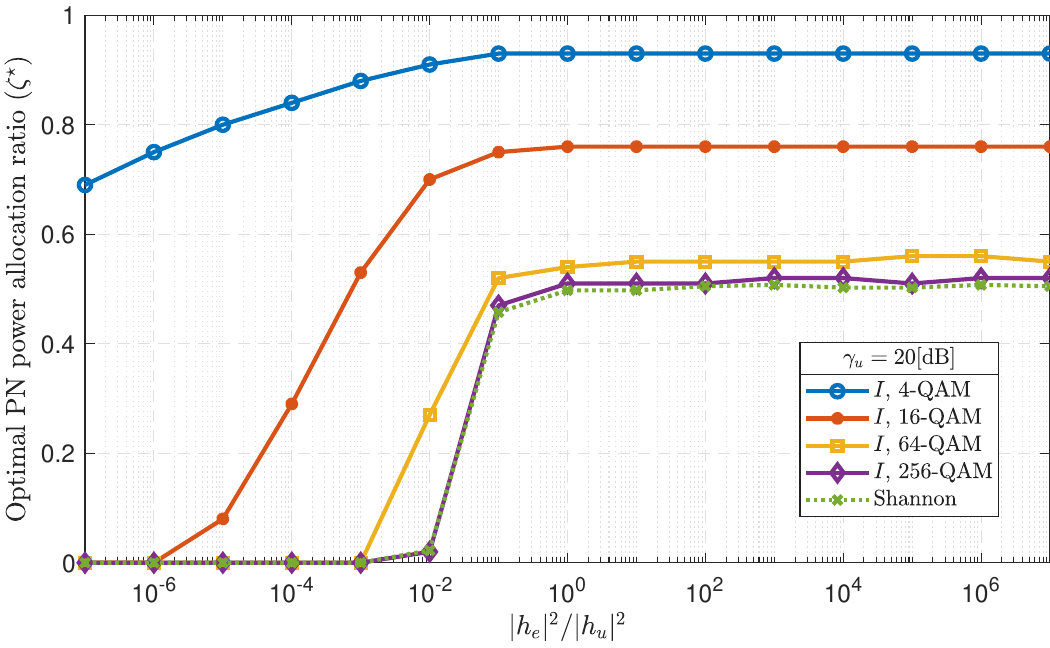}
    \caption{Optimal PN power allocation ratio (\(\zeta^\star\)) versus \(|h_e|^2/|h_u|^2\) for 4, 16, 64 and 256-QAM modulation orders at \(\gamma_u=20\)[dB].}
    \label{fig:Opt_zeta_vs_h_ratio_20dB_M_QAM}
\end{figure}

\section{Conclusions}\label{Sec_Conclusions}
Superimposing a PN signal onto the user’s transmission effectively protects information against eavesdropping, at the cost of allocating part of the transmit power to artificial noise. In many applications, secrecy is prioritized over maximizing the information rate. With finite-alphabet modulation such as $M$-QAM, this power cost causes little performance loss, especially at high SNR, because the achievable rate saturates for each modulation order. In this work, secrecy was evaluated using mutual information, and system performance was analyzed for various channel-quality ratios between the legitimate user and the eavesdropper. The results show that, for a given $M$-QAM order and sufficiently high SNR, near-perfect secrecy is achievable, unlike in systems based on Shannon capacity with unconstrained Gaussian signaling.

Future research will focus on extending the analysis to scenarios involving multiple legitimate users and multiple eavesdroppers. Another promising direction is the design of optimized PN signals that maximize secrecy performance when superimposed on the user’s transmission.

\bibliographystyle{IEEEtran}
\bibliography{Biblio,IEEEabrv}

\end{document}